\documentclass[a4paper,12pt]{article}
\usepackage{amsmath,amscd}
\usepackage{amssymb}
\usepackage{amsmath}
\usepackage{amsfonts}
\usepackage{color}
\usepackage{mathrsfs, graphicx}
\DeclareMathAccent{\wtilde}{\mathord}{largesymbols}{"65}
\DeclareMathAccent{\what}{\mathord}{largesymbols}{"62}

\def\Ad{{\rm Ad}}
\def\ad{{\rm ad}}
\newcommand\cA{{\mathcal A}}
\newcommand\cB{{\mathcal B}}
\newcommand\cS{{\mathcal S}}
\newcommand\cU{{\mathcal U}}
\newcommand\cV{{\mathcal V}}

\newcommand\cI{{\mathcal I}}
\newcommand\cT{{\mathcal T}}
\newcommand\cN{{\mathcal N}}
\newcommand\cW{{\mathcal W}}
\newcommand\Z{{\mathbb Z}}

\newcommand\C{{\mathbb C}}

\newcommand\cR{{\mathcal R}}

\newcommand\cJ{{\mathcal J}}

\newcommand\cL{{\mathcal L}}
\newcommand\cM{{\mathcal M}}

\def\bbbc{{\mathbb C}}

\newcommand{\gM}{{\mathfrak{M}}}

\def\im{\operatorname{Im}}

\def\Tr{{\rm Tr}\,}

\def\mat{{\rm Mat}_{2\times 2}(\cR_x)}

\newcommand\bu{{\bf u}}

\newtheorem{Pro}{Proposition}
\newtheorem{Lem}{Lemma}

\newtheorem{Cor}{Corollary}

\title{
	Formal diagonalisation of Lax-Darboux schemes }

\author{A.V. Mikhailov\\
University of Leeds, UK 
}

\begin{document}
\maketitle

\begin{abstract}
We discuss the concept of Lax-Darboux scheme and illustrate it on well known 
examples associated with the Nonlinear Schr\"odinger (NLS) equation. We explore 
the Darboux links of the NLS hierarchy with the hierarchy of Heisenberg 
model, principal chiral field model as well as with differential-difference 
integrable systems (including the Toda lattice and differential-difference 
Heisenberg chain) and integrable partial difference systems. We show that there 
exists a transformation which formally diagonalises all elements of the 
Lax-Darboux scheme simultaneously. It provides us with generating functions of 
local conservation laws for all integrable systems obtained. We discuss the
relations between conservation laws for systems belonging to the Lax-Darboux 
scheme.     
\end{abstract}

\section{introduction}
Often, integrable partial differential, differential-difference and partial 
difference equations can be regarded as parts of an algebraic structure which 
we call a Lax-Darboux scheme. In this context the Lax representations for the 
integrable partial differential equation (PDE) and a hierarchy 
of its symmetries form a Lax structure. Darboux transformations for the 
corresponding Lax operators are automorphisms of this Lax structure resulting 
in a chain of B\"acklund transformations for the  PDE and its symmetries. The 
latter represents an integrable system of differential-difference equations 
(D$\Delta$Es) and its symmetries (often non-evolutionary). The Bianchi 
permutability conditions for  Darboux transformations represent a system of 
partial difference equations (P$\Delta$Es). This system possesses an infinite 
hierarchy of commuting symmetries (the above mentioned  D$\Delta$Es) and thus 
it 
is an integrable partial difference system in its own right. There are many 
journal publications and monographs focusing on certain aspects of this big 
picture \cite{mat91, mr1908706, BobSuris, kmw13, MPW}. In particular, paper 
\cite{kmw13} contains a good collection of Lax-Darboux representations  
recursion operators for integrable differential-difference equations. In this 
paper we discuss 
the ways in which PDEs, D$\Delta$Es and P$\Delta$Es belonging to the same 
Lax-Darboux scheme 
share the same hierarchy of local conservation laws.

The main idea standing behind our theory is a formal diagonalisation of the 
Lax-Darboux scheme. We show that there exists a formal (i.e. in the form of a 
formal series) gauge transformation which simultaneously diagonalises (or 
brings to a block-diagonal form) the Lax operators of the Lax structure and 
Darboux matrices associated with Darboux transformations. It provides us with a 
regular method for recursive derivation of a hierarchy of local conservation 
laws for the nonlinear differential and difference systems associated with the 
Lax-Darboux scheme.

The method of formal diagonalisation of differential operators can be found in 
the
classical literature concerning asymptotic expansion \cite{vazov}. In 
application to Lax representations 
for partial differential equations and recursive derivation of the hierarchy of 
local conservation laws, there is a neat and instructive exposition of the 
method \cite{mr86h:58071}.  It has 
been successfully used in the Symmetry Approach to classify of integrable 
partial differential equations \cite{mr87h:35312}. Here we extend the method to 
Darboux 
transformations and in this way to differential-difference and partial 
difference integrable systems. We shall explain the method using the 
Lax-Darboux scheme associated with the Nonlinear Schr\"odinger equation. Its 
generalisations to other Lax-Darboux schemes is rather straightforward. 
This paper is based on a lecture courses given by the author in the Bashkir 
State University (Ufa, 2012) and as a part of MAGIC course on Integrable 
systems (UK, 2014), a number of conference talks (Ufa, October 
2012; Moscow, November 2012 \cite{Mik12};  Cambridge, July 2013 
\cite{Mik13}) where the concept of Lax-Darbiux scheme and  formal 
diagonalisation approach were originally presented. This method has proven to 
be useful in a many applications(see for example \cite{MPW, 
habyang,GMY14}).   
  
\section{Lax-Darboux scheme for the Nonlinear Schr\"odinger equation}
  
In this paper we consider Lax integrable equations, i.e. equations which can be 
integrated using the Inverse Spectral Transform method. For such equations we 
can build up a Lax-Darboux scheme with the following objects: 
\begin{itemize}
 \item the {\sl Lax structure}, which is  an infinite 
sequence of Lax operators whose commutativity conditions are equivalent to the 
equation and a
hierarchy of commuting symmetries;
\item {\sl Darboux transformations}, which are automorphisms of the Lax 
structure;
\item {\sl B\"acklund transformations}, which are follow from the 
compatibility conditions of the Lax structure and Darboux transformation. They 
can  be regarded 
as {\sl integrable differential-difference systems};
\item the conditions of Bianchi permutability for the Darboux transformations, 
which lead to systems of {\sl integrable partial difference equations};
\item Ajacent Lax structures associated with a Darboux transformation which 
lead to adjacent symmetries of these  differential-difference and partial 
difference 
equations and are integrable differential-difference equations in their own 
right.
\end{itemize}
In this section we would like to give explicit representations of all 
listed 
above objects in the case of the Nonlinear Schr\"odinger (NLS) equation.
  
\subsection{Lax structure of the Nonlinear Schr\"odinger Equation.}  
  
The Nonlinear Schr\"odinger Equation  is a system of two partial 
differential equations
\begin{equation}\label{nls}
×2p_t=p_{xx}-8p^2q,\qquad
 2q_t=-q_{xx}+8q^2p
\end{equation}
where $x,t$ are independent variables. In the literature the term Nonlinear 
Schr\"odinger Equation usually stands for one complex equation of the form 
\[
 iq_t=q_{xx}\pm 2|q|^2q,
\]
which can be obtained from (\ref{nls}) after a change of variables 
$t\to 2 it,\ x\to 2 ix$ and reduction $p=\mp q^*$. In this paper 
we shall use equation 
(\ref{nls}) to illustrate the method, since 
the reduction condition would add some inessential technicalities.

It has been shown by Zakharov and Shabat \cite{mr53:9966} that the system of 
equations (\ref{nls}) is equivalent to the  commutativity condition
$[L(p,q),A(p,q)]=0$ for two linear differential operators 
\begin{equation}\label{LA}
 L(p,q)=D_x-U,\quad  A(p,q)=D_{t}-V,
\end{equation}
where $D_x$ and $ D_t$ are operators of differentiation in $x$ and $t$ 
respectively,
\begin{equation}\label{UJV}
U=\lambda J+ \left(\begin{array}{rr}
        0& 2p\\2q&0
       \end{array}\right),\ J=\left(\begin{array}{rr}
        1& 0\\0&-1
       \end{array}\right), \ V=\lambda U-\left(\begin{array}{rr}
        2pq& -p_x\\q_x&-2pq
       \end{array}\right),
\end{equation}
and $\lambda$ is a spectral parameter. Linear operators $L(p,q),\  A(p,q)$ form 
a Lax pair (or a Lax representation) for equation (\ref{nls}).

The NLS system (\ref{nls}) admits an infinite hierarchy of commuting symmetries 
\begin{equation}\label{nlshierarchy}
 p_{t_k}=f^k,\quad q_{t_k}=g^k,\qquad k=0,1,2,\ldots
\end{equation}
where
\[\begin{array}{ll}
  f^0=2p,\qquad &g^0=-2q,\\
  f^1=p_x,\qquad &g^1=q_x,\\
  f^2=\dfrac{1}{2}p_{xx}-4p^2q,\qquad &g^2=-\dfrac{1}{2}q_{xx}+4q^2p\qquad
{\mbox{NLS}},\\
  f^3=\dfrac{1}{4}p_{xxx}-6pqp_x,\qquad &g^3=\dfrac{1}{4}q_{xx}-6qpq_x, 
\end{array}\]\[
  \begin{array}{l}
    f^4=\dfrac{1}{8}p_{xxxx}-4pqp_{xx}-3qp_x^2-2pp_xq_x-p^2q_{xx}+12p^3q^2,\\
    g^4=\dfrac{1}{8}q_{xxxx}-4qpq_{xx}-3pq_x^2-2qq_xp_x-q^2p_{xx}+12q^3p^2,\ 
\ldots
   \end{array}
\]
All functions
$f^k,g^k\in [\bbbc,p,q; D_x]$ are differential polynomials over the field
$\bbbc$ of variables 
$p^{(0)}=p,\ q^{(0)}=q$ and their $x$-derivatives
$p^{(1)}=p_x,\ q^{(1)}=q_x,\ p^{(2)}=p_{xx},\ q^{(2)}=q_{xx},\ \ldots$ and 
\[
 D_{x}=\sum_{n=0}^\infty p^{(n+1)}\frac{\partial}{\partial
p^{(n)}}+q^{(n+1)}\frac{\partial}{\partial q^{(n)}}.
\] 
By  (generalised) symmetries of the NLS equation (\ref{nls}) we understand
derivations
\[
 D_{t_k}=\sum_{n=0}^\infty D_x^n(f^k)\frac{\partial}{\partial
p^{(n)}}+D_x^n(g^k)\frac{\partial}{\partial q^{(n)}}
\]
commuting with  $D_t=D_{t_2}$ and $D_x$
\[ 
 [D_{t_2},D_{t_k}]=0,\qquad [D_x,D_{t_k}]=0. 
\]
Symmetries are called commuting if the corresponding derivations commute. It
is sufficient to verify that
$D_{t_n}(f_m)=D_{t_m}(f_n)$ and $ D_{t_n}(g_m)=D_{t_m}(g_n)$. Motivations and 
general definition of symmetries one can find in 
\cite{ mr93b:58070, integrability}.

Each symmetry from this hierarchy has a Lax representation 
\begin{equation}\label{LAlaxstr}
 p_{t_k}=f^k,\quad q_{t_k}=g^k\ \Leftrightarrow\ [L(p,q),A^k(p,q)]=0
\end{equation}
with the same 
operator $L(p,q)=D_x-U$ and $A^k(p,q)=D_{t_k}-V^k$ where matrices $V^k$  can 
be found recursively starting from $V^0=J$ and for $k\ge 1$
\begin{equation}\label{recV}
 V^{k+1}=\lambda 
V^k-\frac{1}{2}D_x(V^k)J-\frac{1}{2}[V^k,U]J-\frac{1}{2}D_x^{-1}{\rm Tr} 
(UD_x(V^k))J.
\end{equation}
Here $D_x^{-1}$ stands for integration in $x$. It can be rigorously proven 
that ${\rm Tr} 
(UD_x(V^k))\in{\rm Im}\, D_x$ for any $k$, thus the integral can be evaluated 
and the result belongs to the differential ring $[\C;p,q;D_x]$. The constants 
of integration can be chosen arbitrary, or fixed by the condition 
$V^k|_{p(x)=q(x)=0}=\lambda^k J$. In the latter case
\[
V^0=J,\  V^1=U,\ V^2=V,\ V^3=\lambda V^2+\frac{1}{2}\left(\begin{array}{cc}
2pq_x-2qp_x& p_{xx}-8p^2 q\\
 q_{xx}-8q^2 p  &2qp_x-   2pq_x                                           ×
                                               \end{array}\right) 
,\ldots \ .\]
Recursion relation (\ref{recV}) can be simplified and represented in the form
\[
 V^0=J,\qquad V^{n+1}=\lambda V^n +B_n,\quad n=0,1,\ldots
\]
where the $\lambda$--independent matrices $B_n$ can be found recursively
\[
B_0=\left(\begin{array}{rr}
        0& 2p\\2q&0
       \end{array}\right),\quad 
B_{k+1}=\frac{J}{2}\left(D_x( B_k)+[B_k,B_0]-D_x^{-1}{\rm 
Tr} 
(B_0D_x(B_k))\right) .
\]

The set of Lax operators $\{L(p,q),A^k(p,q),\ k\in\Z_{\ge 0}\}$ and 
corresponding compatible partial differential equations (commuting symmetries) 
$\{(f^k,g^k),\  k\in\Z_{\ge 0}\}$ form the Lax structure for the Nonlinear 
Schr\"odinger equation (\ref{nls}).

\subsection{Darboux  and B\"acklund transformations for  NLS }
\label{subsec12}
 
Since all linear differential operators $\{L(p,q),A^k(p,q),\ k\in\Z_{\ge 0}\}$ 
commute with each other, there exists a common fundamental solution $\Psi$ of 
the linear problems
\begin{equation}\label{lax0}
 L(p,q)\Psi=0,\qquad A^k(p,q)\Psi=0.
\end{equation}
We shall study a transformation $\cS$ of a 
fundamental solution 
\begin{equation}\label{SPsi}
 \cS:\ \Psi\mapsto\overline{\Psi}=M\Psi,\qquad \det\,M\not\equiv 0
\end{equation}
such that the matrix function $\overline{\Psi}$ is a fundamental solution of
the 
linear problems 
\begin{equation}\label{lax1}
  L(\bar p,\bar q)\bar \Psi=0,\qquad A^k(\bar p,\bar q)\bar \Psi=0.
\end{equation}
with new ``updated potentials''  $\bar p, \bar q$. In the literature this type 
of transformation is often referred to as a Darboux transformation and the 
matrix 
$M$ is 
called the Darboux matrix. We shall assume that a Darboux matrix $M$ is a 
rational  function of the spectral parameter $\lambda$, whose entry may depend 
on $p,q,\bar p, \bar q$ and  may also depend on some auxiliary function(s) 
$h$  or constant parameter(s) (examples will be given later in this section). 
The given description of a Darboux transformation can be cast into 
a rigorous definition using
elements of differential-difference ring theory. We wish to avoid the 
introduction of these concepts
at the present time to make the paper accessible to a wider community.

It follows from (\ref{lax0}), (\ref{SPsi}), (\ref{lax1}) that transformation 
$\cS$ can be extended to the set of Lax operators:
\begin{equation}\label{LAaut}
  \begin{array}{l}
   \cS: L(p,q)\mapsto L(\bar p,\bar q) =ML(p,q)M^{-1},\\  \\
   \cS: A^k(p,q)\mapsto A^k(\bar p,\bar q)=M A^k(p,q)M^{-1}, \quad k=0,1,\ldots 
\ .
  \end{array}
\end{equation}
Let us show it for the first equation. Indeed, we have 
\[
 D_x\bar \Psi=\cS (U) \bar\Psi \Rightarrow D_x (M \Psi)=\cS (U) M\Psi 
\Rightarrow D_x(M)-\cS (U) M+MU=0
\]
and thus $L(\bar p,\bar q)M -ML(p,q)=0$.
Here we use notation $\cS (U)$ for matrix $U$ (\ref{UJV}) in which 
variables $p,q$ are replaced by $\bar p,\bar q$. Equations (\ref{LAaut}) are 
equivalent to a compatible system of equations for $M$ 
\begin{eqnarray}
&& D_x(M)=\cS (U)M-MU,\label{backlundx}\\
&& D_{t_k}(M)=\cS (V^k)M-MV^k.\label{backlundtk}.
\end{eqnarray}

It follows from (\ref{LAaut}) that a Darboux transformation $\cS$ can be 
regarded as
an automorphism  of the Lax structure. It maps the
 set of commuting Lax operators  into 
another commuting set  
\[
 \cS:  \{L(p,q),A^k(p,q);\ k\in\Z_{\ge 0}\}\mapsto \{L(\bar p,\bar q),\ 
A^k(\bar p,\bar q);\ k\in\Z_{\ge 
0}\},
\]
and results in a B\"acklund transformation which transforms  a solution $p,q$ 
of 
the  NLS hierarchy into a new solution 
$$\cS:(p,q)\mapsto (\bar p,\bar q).$$
Equations (\ref{backlundx}), (\ref{backlundtk}) follow from the
conditions 
that the map $\cS$ and derivations $D_x, D_{t_k}$ commute 
\[ \cS L=MLM^{-1}:=\Ad_M L,\quad \cS A^k=MA^kM^{-1}:=\Ad_M A^k,\qquad 
k=0,1,\ldots\ .\]

Darboux transformations are obviously invertible and a composition of two 
Darboux transformations $\cS_1,\cS_2$ with Darboux matrices $M_1,M_2$
\[
 \cS_2\circ \cS_1: \Psi\mapsto \cS_1 (M_2)M_1\Psi
\]
is a Darboux transformation. There is a 
problem to describe all {\sl elementary} Darboux 
transformations for a given Lax structure, such that any 
rational Darboux transformation can be represented as a composition of the 
elementary ones. In the cases of the Lax structure corresponding to the 
Korteweg-de Vries equation and the Nonlinear Schr\"odinger equation the 
solution of this problem can be found in \cite{adler_dis}. In particular, it 
has 
been shown  that any Darboux transformation of the Lax 
operators $L,A$ (\ref{LA}) can be represented as a composition of elementary 
Darboux transformations $\cJ_\beta, \cS_{\alpha}, \cT_h $ with matrices 
\begin{eqnarray} 
 &&
\cJ_\beta:
J_\beta=\mbox{diag}(\beta,\beta^{-1}),\\ \label{Malpha}
&&\cS_{\alpha}:             
M_\alpha=\left(\begin{array}{cc}
                               \lambda\!+\!p\cS_{\alpha}(q)\!-\!\alpha&p\\
\cS_{\alpha}(q)&1
                              \end{array}\right),\\    \label{Nh}            
&&\cT_h:  N_h=\left(\begin{array}{cc}
                               \lambda\!+\!h&p\\\cT_h(q)&0
                              \end{array}\right),
\end{eqnarray}
and their inverse transformations with a certain choice of constant complex 
parameters $\beta, \alpha$.

It follows from (\ref{backlundx}) that
\begin{eqnarray}\label{Jeq}
&& \cJ_\beta L=\Ad_{J_\beta}L\ \Leftrightarrow \ \cJ_\beta(p)=\beta^2 p,\
\cJ_\beta(q)=\beta^{-2} q;\\&&\nonumber\\ \label{Meq}
&& \cS_\alpha L=\Ad_{M_\alpha}L\  \Leftrightarrow \ \left\{\begin{array}{l}
p_x=2\cS_\alpha(p)-2p^2\cS_\alpha(q)+2\alpha p, \\ \\
\cS_\alpha (q_{x})=-2 q +2p\cS_\alpha (q^2)-2\alpha\cS_\alpha ( q);
                                            \end{array}
\right.
\\&& \phantom{{X}}\nonumber \\  \label{Neq}
&& \cT_h L=\Ad_{N_h}L\  \Leftrightarrow \ \left\{\begin{array}{l}p_x=-2hp,\
h_x=2(\cT_h-1)(pq),\\ \\ p\cT_h(q)=1.\end{array}
\right.
\end{eqnarray}

The first map (\ref{Jeq})
\[
  \cT_\beta: (p,q)\mapsto      (\beta^2 p,\beta^{-2}q)      
              \]
is nothing but a point symmetry of the NLS equation (\ref{nls}).

Equation (\ref{Meq}) is a B\"acklund transformation (the $x$ part of the 
B\"acklund transformation) for the NLS equation (\ref{nls}). Starting from a 
solution $p,q$ of the NLS we can find a new solution $(p_1,q_1)=(\cS_\alpha 
(p),\cS_\alpha (q))$ by solving a Riccati equation for $q_1$
\[
 q_{1,x}=-2 q +2pq_1^2-2\alpha q_1
\]
and then $p_1=p^2q_1-2\alpha p-p_x$. Equations  (\ref{Meq}) can be regarded as 
an integrable system of differential--difference equations (D$\Delta$Es). In 
variables
\begin{equation}\label{pqn}
p_n =\cS_\alpha^n(p),\ q_n=\cS_\alpha^n(q),\ 
\alpha_n=\cS_\alpha^n(\alpha)
\end{equation}
it takes the form \cite{mrt94}
\begin{equation}\label{meq}\begin{array}{lll}
     p_{n,x}&=&2p_{n+1}-2p_n^2q_{n+1}+2\alpha_n p_n, \\
q_{n,x}&=&-2q_{n-1}+2p_{n-1}q_n^2-2\alpha_{n-1} q_n,
\end{array}\qquad n\in\Z,\ \ \alpha_n\in\C.
\end{equation}
In order to simplify notations we often shall omit $n$ in the lower index 
for functions depending on the point of the lattice replacing $p_{n\pm 1}$ by 
$p_{\pm 1}$, etc.
System (\ref{meq}) has a Lax-Darboux representation (often called a 
semi-discrete Lax representation)
\[
 L\Psi=0,\qquad \cS_\alpha (\Psi)=M_\alpha \Psi.
\]
Its compatibility condition  (\ref{backlundx}) is equivalent 
to (\ref{meq}). It has an infinite hierarchy of commuting symmetries following 
from the conditions $\cS_\alpha A^k=\Ad_{M_\alpha}A^k$ and  local conservation 
laws. The 
latter will be shown in the next section.
 
Automorphism $\cT_h$ gives the explicit map for solutions of the NLS:
\begin{equation}\label{todamap}
\cT_h( p)=p^2q-\frac{p}{4}\left(\frac{p_x}{p}\right)_x,\qquad
\cT_h(q)=\frac{1}{p}.
\end{equation}
In variables $p=\exp\phi,\  \phi_{k}=\cT_h^k(\phi),\ h_k=\cT_h^k(h)$ 
it can be written as 
\begin{equation} 
 \phi_{x}=-2h,\qquad 
  h_{x}=2\exp(\phi_1-\phi)-2\exp(\phi-\phi_{-1})
\label{todax}
\end{equation}
and after elimination of $h$ it takes the form of the Toda lattice:
\[\phi_{xx}=4\exp(\phi-\phi_{-1})-4\exp(\phi_{1}-\phi). 
\]
The D$\Delta$Es which follow from 
$\cT_h A^k=\Ad_{N_h}A^k$ are
symmetries of (\ref{todax}). For example  for $k=0,1,2,3$ 
we obtain
\[ \begin{array}{l}
\left\{ \begin{array}{l}
  \phi_{t_0}=2,\\
  h_{t_0}=0;
 \end{array}\right.\qquad 
\left\{  \begin{array}{l}
  ×\phi_{t_1}=-2h,\\
  h_{t_1}=2(\cT_h-1)\exp(\phi-\phi_{-1});
\end{array}\right.\\ \\
\left\{\begin{array}{l}
  \phi_{t_2}=2h^2-2(\cT_h+1)\exp(\phi-\phi_{-1}),\\ 
  h_{t_2}=-2 (\cT_h-1)( \exp(\phi-\phi_{-1})(h_{-1}+h));
\end{array}\right.
\\ \\
\left\{ \begin{array}{l}
\phi_{t_3}=2\exp(\phi-\phi_{-1})(2h+h_{-1})+2\exp(\phi_1-\phi)(2h+h_{1})-2h^3,
\\ \\
  h_{t_3}=2
(\cT_h-1)((h_{-1}^2+h_{-1}h+h^2)\exp(\phi-\phi_{-1})+\\
\qquad \ \  \exp(2\phi-2\phi_{-1})+(\cT_h+1)\exp(\phi-\phi_{-2})).  
\end{array}\right.\
   \end{array}
\]
To define the explicit map   (\ref{todamap}) we have to consider the
localisation of the ring with respect to the element $p^{-1}$. Then we
introduced the exponential function in order to transform the system in the
standard well known form of the Toda lattice.
  
\subsection{Bianchi commutativity of Darboux maps and integrable P$\Delta$Es}

Let us impose the condition that automorphisms corresponding to two Darboux
maps $\cS_\alpha$ and $\cS_\beta$ commute
\begin{equation}\label{mamb}
 [\cS_\alpha,\cS_\beta]=0\quad \Rightarrow\quad
\cS_\alpha(M_\beta)M_\alpha-\cS_\beta(M_\alpha)M_\beta=0 .
\end{equation}
It leads to the Bianchi lattice 
which can be regarded as a system of partial difference equations (P$\Delta$E) 
on  $\Z^2$.
\[ 
 \left\{
\begin{array}{l}
  (\cS_\beta(p)-\cS_\alpha(p))(1+p\cS_\beta\cS_\alpha(q))-(\alpha-\beta)p=0\\ \\
 (\cS_\beta(q)-\cS_\alpha(q))(1+p\cS_\beta\cS_\alpha(q))+(\alpha-\beta)\cS_\beta
\cS_\alpha(q)=0
                                                     \end{array}\right.
\]
Denoting $p_{nm}=\cS_\alpha^n\cS_\beta^m(p), q_{nm}=\cS_\alpha^n\cS_\beta^m(q)$
we get a quadrilateral system of equations:
\begin{equation} \label{dnls}
p_{01}\,=\,p_{10}\,+\,\frac{\alpha-\beta}{1+ p\,q_{11}}\,p\,,\
q_{01}\,=\,q_{10}\,-\,\frac{\alpha-\beta}{1+ p\,q_{11}}\,q_{11}\,. \
\end{equation}
D$\Delta$Es, which follow from conditions 
\begin{equation}\label{symxtk}
\cS_\alpha L=\Ad_{M_\alpha}L,\ \cS_\alpha A^k=\Ad_{M_\alpha}A^k
\end{equation}
are generalised symmetries of (\ref{dnls}). Symmetries corresponding to
conditions $\cS_\beta L=\Ad_{M_\beta}L,\  \cS_\beta A^k=\Ad_{M_\beta}A^k$ are 
equivalent to 
(\ref{symxtk}) modulo system (\ref{dnls}).

Similarly, the condition $ [\cS_\alpha,\cT_h]=0$ leads to 
\begin{equation}\label{manh}
 \cT_h(M_\alpha)N_h=\cS_\alpha(N_h)M_\alpha,
\end{equation}
which is equivalent to the fully discrete 
Toda lattice
\begin{equation}\label{dtoda}
{\rm{e}}^{\phi_{10}-\phi} - {\rm{e}}^{\phi-\phi_{-10}} +
{\rm{e}}^{\phi_{-11}-\phi} - {\rm{e}}^{\phi-\phi_{1,-1}} + \alpha
-\alpha_{-1}=0,
\end{equation}
and
$$
h={\rm{e}}^{\phi-\phi_{1,-1}}-{\rm{e}}^{\phi_{10}-\phi}-\alpha,\qquad
$$
in the variables $\phi_{pq}=\cT_h^p\cS_\alpha^q \log p $  and 
$\cT_h^k\alpha=\alpha, \cS^k\alpha=\alpha_k, \alpha_k\in\C  $. The discrete 
Toda lattice is a difference equation which is defined on a 5--points stencil.

Equations  which follow from conditions 
\begin{equation}
 \label{symtoda}
\cS_\alpha L=\Ad_{M_\alpha}L,\ \cS_\alpha A^k=\Ad_{M_\alpha}A^k
\end{equation}
are
symmetries of the discrete Toda lattice. For example $\cS_\alpha 
L=\Ad_{M_\alpha}L$ results in
 $$ \phi_x=\,- 2({\rm{e}}^{\phi-\phi_{1,-1}} -
{\rm{e}}^{\phi_{10}-\phi} - \alpha).$$
Symmetries corresponding to
conditions $\cT_h L=\Ad_{N_h}L,\  \cT_h A^k=\Ad_{N_h}A^k$ are 
equivalent to 
(\ref{symtoda}) modulo system (\ref{dtoda}).

\subsection{Adjacent Lax structure}\label{alternative}

In Section \ref{subsec12} we discussed the problem to find all elementary 
Darboux matrices corresponding to a given Lax structure. There is another 
interesting and important problem to find all possible Lax structures 
associated with a given Darboux matrix. In this section we show that the
Darboux 
matrix $M_\alpha$ (\ref{Malpha}) corresponding to Lax operators (\ref{LA})  
admits an alternative Lax structure with operators $B^k=D_{ y_k}-W^k$. We 
shall assume that $\cS_{\alpha}(\alpha)=\alpha$, i.e. the constant $\alpha$ 
does not depend on the vertex of the lattice.
  
We notice that the determinant of the Darboux matrix
 \[M_\alpha=\left(\begin{array}{cc}
                               \lambda\!+\!p\cS_\alpha(q)\!-\!\alpha&p\\
\cS_\alpha(q)&1
                              \end{array}\right)\]
is $\lambda-\alpha$. Thus at $\lambda=\alpha$ $M^0_\alpha= 
M_\alpha\lvert_{\lambda=\alpha}$ has rank $1$ and can be represented by a 
bi-vector
\[M^0_\alpha=\left(\begin{array}{c}
p\\1                                           ×
                                          \end{array}\right)\cdot \left( 
\cS_\alpha(q),\ 1\right).
\]

Let us search for a Lax operator $B^1_\alpha=D_{ y}-W^1_\alpha$ 
with a matrix $W^1_\alpha$ having a simple pole with a residue of rank 1 at  
$\lambda=\alpha$ and vanishing at $\lambda=\infty$
\[
 W^1_\alpha=\frac{\cW^1_\alpha}{\lambda-\alpha}.
\]
It follows from 
$\cS_\alpha B^1_\alpha=\Ad_{M_\alpha}B^1_\alpha$ that
\begin{equation}
\label{W1eq}
D_{ y}(M_\alpha)=\cS_\alpha(W^1_\alpha)M_\alpha-M_\alpha W^1_\alpha.
\end{equation}
Taking the residue at $\lambda=\alpha$ we get equation 
\[
 S_\alpha (\cW^1_\alpha)M^0_\alpha=M^0_\alpha \cW^1_\alpha
\]
which has a unique (up to a scalar constant factor $\gamma$) solution 
\[
\cW^1_\alpha=\frac{\gamma}{1+\cS_\alpha^{-1}(p)\cS_\alpha(q)}\left(\begin{array}
{c}
\cS_\alpha^{-1}(p)\\1                                           ×
                                          \end{array}\right)\cdot \left( 
\cS_\alpha(q),\ 1\right).
\]
In what follows we set $\gamma=1$. Thus 
\[ 
W^1_\alpha=\frac{1}{(\lambda-\alpha)(1+\cS_\alpha^{-1}(p)\cS_\alpha(q))}
\left(\begin{array}{cc}
\cS_\alpha^{-1}(p)\cS_\alpha(q)&\cS_\alpha^{-1}(p)\\
\cS_\alpha(q)&1
                              \end{array}\right).
\]
Entries of $W^1_\alpha$ are not from the differential-difference polynomial
ring and  localisation of the ring by  the element
$(1+\cS_\alpha^{-1}(p)\cS_\alpha(q))^{-1}$ is required.

With this $W^1_\alpha$ equation (\ref{W1eq}) is equivalent to the following 
evolutionary system of integrable differential-difference equations
\begin{equation}\label{dH}
p_{ y}=-\frac{p_{-1}}{1+p_{-1}q_1}, \
q_{ y}=\frac{q_1}{1+p_{-1}(p)q_1}. 
\end{equation}
Here we use notations introduced in (\ref{pqn}).
System (\ref{dH}) is a new symmetry of differential-difference system
(\ref{meq}) as well as of partial-difference systems (\ref{dnls}),(\ref{dtoda}).

It can be shown that there is an infinite hierarchy of commuting operators 
\begin{equation}\label{Balphak}
B_\alpha^k=D_{ y_k}-W_\alpha^k,\qquad 
 W^{k+1}_\alpha=\frac{1}{\lambda-\alpha}(W^k_\alpha+C^k)
\end{equation}
and $C^k$ is a $\lambda$-independent matrix. Using condition 
$[B^1_\alpha,B^2_\alpha]=0$  we can find that 
\[
 C^1=\frac{1}{(1+p_{-1}q_1)^2}\left(\begin{array}{cc}
 p_{-1}q_{1, y}-p_{-1, y}q_1& -p_{-1, y}-p_{-1}^2q_{1, y}\\
 q_{1, y}+q_{-1}^2p_{-1, y}&p_{-1, y}q_1-p_{-1}q_{1, y}
                                    \end{array}\right).
\]
For matrices $W^k_\alpha$ there exists a recursion (similar to (\ref{recV}))
which enables one to find the infinite hierarchy of operators $B^k_\alpha$ 
recursively.  Operators $B_\alpha^k,\ k=1,2,\ldots$ form the {\sl adjacent} Lax 
structure.

The 
partial differential equation which is equivalent to the condition  
$[B^1_\alpha,B^2_\alpha]=0$ is of the form 
\begin{equation}\label{1H}
 \begin{array}{l}
( p_{-1})_{  y_2}=- ( p_{-1})_{ y y}+\dfrac{2q_1 
((p_{-1})_{ y})^2}{1+p_{-1}q_1},\\ \\
( q_{1})_{  y_2}= ( q_{1})_{ y y}-\dfrac{2p_{-1} 
((q_{1})_{ y})^2}{1+p_{-1}q_1}.
 \end{array}
\end{equation}
System (\ref{dH}) is a B\"acklund transformation for (\ref{1H}).
Equation (\ref{1H}) is well known, it is a Heisenberg model for ferromagnets 
\cite{ZakTak79}
\[
 {\bf S}_{\tau}={\bf S}\times {\bf S}_{ y y},\qquad {\bf S}^2=1
\]
after the change of variable $ y_2=i\tau$ and  stereographic projection 
\[
 {\bf S}=\left(
\frac{p_{-1}+q_1}{1+p_{-1}q_1} ,\ 
i\frac{q_1-p_{-1}}{1+p_{-1}q_1} ,\ \frac{ p_{-1}q_1-1}{1+p_{-1}q_1}\right).
\]

We can use equation (\ref{dH}) to eliminate $ y$ derivatives from the Lax 
operator $B^2_\alpha$ and partial differential equation (\ref{1H}). The 
latter will take the form of a differential-difference system
\begin{equation}\label{dH2}
 \begin{array}{l}
p_{ y_2}=\dfrac{p_{-1}^2q_2(1+p_{-2}q)-p_{-2}(1+pq_2)}{(1+p_{-2}
q)(1+p_{-1}q_1)^2(1+pq_2)},\\ \\
q_{ y_2}=\dfrac{q_{2}(1+p_{-2}q)-q_{1}^2p_{-2}(1+pq_2)}{(1+p_{-2}
q)(1+p_{-1}q_1)^2(1+pq_2)}.
 \end{array}
\end{equation}
Equation (\ref{dH2}) is a B\"acklund transformation for (\ref{1H}) and a 
symmetry of systems (\ref{dH}), (\ref{meq}), (\ref{dnls}), (\ref{dtoda}).

The system of partial difference equations (\ref{dnls}) is equivalent  to the 
commutativity of Darboux transformations $[\cS_\alpha,\cS_\beta]=0$, 
corresponding to Darboux matrices $M_\alpha$ and $M_\beta$, with distinct 
values of the parameters $\alpha$ and $\beta$. With these matrices we associate 
two Lax operators
\begin{equation}\label{cAalpha}
 B_\alpha=D_{ y}-\frac{W^1_\alpha}{\lambda-\alpha},\quad 
B_\beta=D_{z}-\frac{W^1_\beta}{\lambda-\beta}, \qquad \alpha\ne\beta ,
\end{equation}
which coincide with the Lax pair for the principal chiral field model 
\cite{mr80c:81115}. The compatibility condition $[B_\alpha,B_\beta]=0$ 
leads to the system
\[
 (W^1_\beta)_ y = \frac{[W^1_\alpha,W^1_\beta]}{\beta-\alpha},\qquad 
(W^1_\alpha)_z = \frac{[W^1_\alpha,W^1_\beta]}{\beta-\alpha},
\]
which in  variables $p_{nm}=\cS_\alpha^n\cS_\beta^m(p), 
q_{nm}=\cS_\alpha^n\cS_\beta^m(q)$can be written as
\begin{equation}\label{chiralfield}
 \begin{array}{l}
(p_{-1,0})_z=\dfrac{(p_{0,-1}-p_{-1,0})(1+p_{-1,0}q_{0,1})}{
(\alpha-\beta)(1+p _{0,-1}q_{0,1})},\\ \\
(q_{1,0})_z=\dfrac{(q_{1,0}-q_{0,1})(1+p_{0,-1}q_{1,0})}{
(\alpha-\beta)(1+p_{0,-1}q_{0,1})},\\ \\
(p_{0,-1})_ y=(p_{-1,0})_z,\qquad 
(q_{0,1})_ y=(q_{1,0})_z.
 \end{array}
\end{equation}

Finally, let us consider the compatibility condition $ [L,B_\alpha]=0$ for 
two linear problems 
\[ L\Psi=0,\qquad B_\alpha\Psi=0 \]
with the original Lax operator  $L$ (\ref{LA}) and operator 
$B_\alpha$(\ref{cAalpha}). 
Vanishing of the commutator at infinity in $\lambda$ is equivalent to  
equations (\ref{dH}). Using equations (\ref{dH}) we can express
$p_{-1}$ and $q_1$ as
\[
 p_{-1}=\dfrac{1+ \sqrt{1+4(p)_y (q)_y}}{2(q)_y},\quad q_{1}=-\dfrac{1+ 
\sqrt{1+4(p)_y (q)_y}}{2(p)_y}.
\]
Of course there is also the second solution with the negative sign at the 
square  root, it can be treated similarly.
Then the compatibility conditions are equivalent to the system of partial 
differential equations
\begin{equation}\label{pqxix}
 (p)_{xy}=2\alpha (p)_y+ 2p\sqrt{1+4(p)_y (q)_y},\quad (q)_{xy}=-2\alpha 
(q)_y+ 2q\sqrt{1+4(p)_y (q)_y}.
\end{equation}
The constant $\alpha$ can be removed by a simple change of variables 
$P=pe^{-2\alpha x},\ Q=qe^{2\alpha x}$. Then the system admits an obvious 
reduction $P=Q$ to a single hyperbolic equation
\[
 (P)_{xy}=P\sqrt{1+4((P)_ y)^2}.
\]
The latter equation is well known in the literature. It can be reduced to the 
sine-Gordon equation by a differential substitution \cite{mr1845643}. The above 
construction provides us with the Lax representation for the system 
(\ref{pqxix}) with the Lax operators $L$ (\ref{LA}) and the second operator 
\begin{equation}\label{bsok}
 B=D_ y-\frac{1}{2(\lambda-\alpha)}\left(\begin{array}{cc}
\sqrt{1+4(p)_y (q)_y}&-2(p)_y\\2(q)_y&-\sqrt{1+4(p)_y (q)_y}                    
        
                                            \end{array}\right).
\end{equation}
Similarly one can eliminate shifts from operators $B_\alpha^k$ to build up the 
Lax structure correponding to (\ref{pqxix}), (\ref{bsok}).

\section{Formal diagonalisation of the Lax-Darboux Scheme}

In this section we show that the Lax operators and corresponding Darboux 
matrices can be simultaneously formally diagonalised. The resulting objects 
will be presented by formal Laurent  expansions at poles of the chosen 
Lax operator. It will enable us to find recursively local conservation laws 
for corresponding partial differential, differential-difference and partial 
difference equations simultaneously. Our aim is to show that for 
evolutionary equations from the same Lax-Darboux scheme with Lax operators
$L,A^k,\ k=0,1,...$ and Darboux maps $\cS_i,\ i=1,2,...$ there is an infinite
sequence of common local conservation laws with densities $\rho_n,r^i_n$ 
and fluxes $\sigma_n^k$. That is,
\begin{enumerate}
 \item PDE's corresponding to the Lax structure  $ [L,A^k]=0$ possess 
conservation lows 
\[
D_{t_k}\rho_n=D_x\sigma_n^k,\qquad k=0,1,\ldots\ .
\]
\item For differential-difference equations originating from the conditions 
$\cS_i L=\Ad_{M_i}L$ and $\cS_i A^k=\Ad_{M_i}A^k$ there are conservation laws 
of 
the form 
\[
 D_x r^i_n=(\cS_i-1)\rho_n,\qquad D_{t_k} r^i_n=(\cS_i-1)\sigma_n^k
\]
with the same $\rho_n$ and $\sigma_n^k$ as above, modulo equation 
$\cS_i L=\Ad_{M_i}L$.
\item For partial difference equations, corresponding to the Bianchi lattice 
$[\cS_i,\cS_j]=0$ the corresponding sequence of the conservation laws are:
\[(\cS_j -1)r^i_n=(\cS_i -1)r^j_n.\]
\end{enumerate}

We shall demonstrate it on the examples 
described in the previous section, associated with the Lax-Darboux scheme 
for the nonlinear Schr\"odinger equation as well as with the adjacent Lax 
structures considered in Section  \ref{alternative}. A generalisation of this 
approach to other Lax-Darboux schemes (or their parts) often is rather 
straightforward and it will be discussed at the end of this Section. 

\subsection{Formal diagonalisation of the Lax structure for NLS
($L,A^k$).}

In the Lax operator $L$ (\ref{LA}) matrix $U$ has a simple pole in $\lambda$ at 
infinity with the coefficient  $J$ which is  diagonal (\ref{UJV}). The 
matrices $V^k$ in operators $A^k=D_{t_k}-V^k$ are differential polynomials in 
variables $p,q$ and their $x$ derivatives with complex coefficients. The 
leading (in $\lambda$) coefficient is also diagonal and is equal to $\lambda^k 
J$. By 
{\sl local} functions (in this case) we shall understand elements of the 
differential polynomial ring  $\cR_x=[\C;p,q;D_x]$.

Let us consider endomorphism $\ad_J$ of the linear space $\gM=\mat$ of
$2\times 
2$ matrices with entries from $\cR_x$ 
\[ \ad_J:\gM\mapsto\gM, \quad \ad_J (a)=Ja-aJ,\quad a\in\gM.
\]
The kernel of $\ad_J$ is the subspace of diagonal matrices, the image space of 
$\ad_J$ is a subspace of off-diagonal matrices. Thus 
\[
 \gM=\gM_\parallel\oplus\gM_\perp,\qquad \gM_\parallel={\rm Ker}\, \ad_J,\ 
\gM_\perp={\rm Im}\,\ad_J.
\]
In $\gM_\perp$ the endomorphism $\ad_J$ is invertible
\[
 \ad_J^{-1}:\gM_\perp\mapsto\gM_\perp, \quad \ad_J^{-1}a=\frac{1}{4}\ad_J 
a,\qquad\forall a\in\gM_\perp.
\]
In the space 
$\gM$ it is convenient to introduce two projectors 
\[
 \pi_\perp=\frac{1}{4}\ad_J^2,\quad \pi_\parallel={id}-\pi_\perp
\]
where ${ id}$ is the identity map. They are projectors on the off-diagonal and 
diagonal part of a matrix respectively
\[
 \pi_\perp \gM=\gM_\perp,\qquad \pi_\parallel\gM=\gM_\parallel .
\]

We shall use a simplified version of the Drinfeld-Sokolov Lemma, which they 
have formulated and proved in a rather general setting \cite{mr86h:58071}. 

\begin{Lem}
 
For linear operator $L$ (\ref{LA}) there exists a unique formal
series 
\begin{equation}\label{Q}
 Q=I+\lambda^{-1}Q_1+\lambda^{-2}Q_2+\lambda^{-3}Q_3+\cdots,\qquad
Q_k\in\gM_\perp
\end{equation}
 such that 
\begin{equation}\label{ccL}
 \cL=Q^{-1}LQ=D_x-\lambda
J-\cU_0-\lambda^{-1}\cU_1-\lambda^{-2}\cU_2-\cdots,\qquad
\cU_k\in\gM_\parallel .
\end{equation}
 The 
coefficients $Q_k$ can be found recursively
\begin{equation}\label{Qk}
 Q_1=-\frac{1}{4} \ad_J U,\qquad  
Q_{k+1}=\frac{1}{4} \ad_J\left(D_x(Q_k)+\sum_{p=1,q=1}^{p+q=k}Q_pUQ_q\right) 
,
\end{equation}
and
\[
\cU_0=0,\qquad \cU_k=UQ_k\ .
\]
\end{Lem}

{\bf Proof.}
 Let us substitute $L,\cL$ and $Q$ in 
\[
 C=LQ-Q\cL=C_0+\lambda^{-1}C_1+\lambda^{-2}C_2+\cdots\ .
\]
The condition that the formal series $C$ should vanish  provides us with
a sequence of equations to determine the coefficients $\cU_k,Q_k$. The 
linear in $\lambda$ term  vanishes automatically. The coefficient at $\lambda^0$ 
is
\[
 C_0=[J,Q_1]+U-\cU_0.
\]
Applying projectors $\pi_\parallel$ and $\pi_\perp$ to $C_0$ we find
that
\[
 \pi_\parallel ([J ,Q_1]+U-\cU_0)=-\cU_0=0,\qquad
\pi_\perp([J ,Q_1]+U-\cU_0)=[J ,Q_1]+U=0.
\]
Thus $\cU_0=0$ and $Q_1=-\frac{1}{4}\ad_J  U$. The coefficient at
$\lambda^{-k}$ is
\[
 C_k=\ad_J  Q_{k+1}-D_x(Q_k)+UQ_k-\cU_k-\sum_{p=1,q=1}^{p+q=k}Q_p\cU_q.
\]
Therefore
\[
 \pi_\parallel(C_k)=UQ_k-\cU_k=0\ \Rightarrow \cU_k=UQ_k,
\]
\[
 \pi_\perp(C_k)=\ad_J 
Q_{k+1}-D_x(Q_k)-\sum_{p=1,q=1}^{p+q=k}Q_p\cU_q=0\ .
\]
Thus the coefficients $Q_n,\cU_n$ can be found recursively:
\[
 Q_1=-\frac{1}{4}\ad_J  U,\ Q_{k+1}=\frac{1}{4} 
\ad_J\left(D_x(Q_k)+\sum_{p=1,q=1}^{p+q=k}Q_pUQ_q\right) ,\ \cU_k=UQ_k.\ \
\]
Note that $Q_k$ and $\cU_k$ are all local, i.e. expressed in terms of 
differential 
polynomials.\hfill $ {\Box}$ 
\medskip

We are going to
show that $Q$ diagonalises the whole Lax-Darboux scheme, i.e. diagonalises the 
operators
$A^k$ and the Darboux matrices $M_\alpha$ (corresponding to $\cS_\alpha$). It 
is well known that all operators $A^k$ in the Lax structure become diagonal and 
lead 
to local conservation laws for the corresponding partial differential equations 
and their symmetries. 
\begin{Pro} Let $[L,A^k]=0$, where 
\[A^k=D_{t_k}-\lambda^k 
J-\lambda^{k-1}V^k_{1-k}-\lambda^{k-2}V^k_{2-k}-\cdots\ .\]
 Then 
 \begin{equation}\label{ccAk}
  \cA^k= Q^{-1}A^k Q =D_{t_k}-\lambda^k 
J-\lambda^{k-1}\cV^k_{1-k}-\lambda^{k-2}\cV^k_{2-k}+\cdots
 \end{equation}
has  diagonal coefficients $ \cV^k_{s}\in\gM_\parallel, \ s=k-1,k-2,\ldots$ .
\end{Pro}
{\bf Proof.} If  $[L,A^k]=0$, then $[\cL,\cA^k]=0$ where $\cL=Q^{-1}L Q$  
(\ref{ccL}) and  $\cA^k=Q^{-1}A^k Q$ (\ref{ccAk}). Using induction we show that 
all coefficients of the formal series $\cA^k$ are diagonal. The leading 
coefficient of the series $\lambda^k J$ is diagonal. Let us assume that 
coefficients $\cV^k_{1-k},\cV^k_{2-k},\ldots , \cV^k_{n-k}$ are diagonal. Then
the leading  term of 
 $\pi_\perp [\cL,\cA^k]$ is equal to $\lambda^{k-n} \ad_J \cV^k_{n+1-k}$. It 
should vanish and thus $\cV^k_{n+1-k}\in \gM_\parallel$.\hfill $\Box$
\begin{Cor}
The following ystems of partial differential equations 
\[
(p_{t_k}=f^k,\ q_{t_k}=g^k)\ \Leftrightarrow\ [L,A^k]=0
\]
have an infinite hierarchy of common conservation laws
\[
 (\cU_n)_{t_k}=D_x(\cV^k_n),\qquad n=1,2,\ldots\ ,
\]
Moreover,   $D_x \cV^k_m=0$, for $m=1-k,2-k,\ldots,1, 0$.
\end{Cor}
{\bf Proof.} From $[L,A^k]=0$ it follows that $[\cL,\cA^k]=0$ which leads to
\[
 \sum_{n=1}^\infty \lambda^{-n}(\cU_n)_{t_k}-\sum_{n=1-k}^\infty 
\lambda^{-n}(\cV^k_n)_x=0.
\]
Vanishing the coefficients at each power $\lambda^n$ proves the 
statement.\hfill $\Box$ 
 
It is easy to show that differential polynomials $\Tr \cU_n\in \im D_x$ and 
thus they correspond to trivial densities. Let us take the matrix entry 
$(\cU_n)_{2,2},\ (\cV^k_n)_{2,2}$ to define
\[
 \rho_n=(\cU_n)_{2,2},\qquad \sigma_n^k=(\cV^k_n)_{2,2}.
\]
It can be shown that the corresponding densities 
are all non-trivial.

{\bf Example.} Taking $L$ corresponding to the NLS (\ref{LA}) we find that 
\[
\begin{array}{ll}
Q_1=\left(\begin{array}{cc}
            0&-p\\q&0
           \end{array}\right), \qquad &
\cU_1=2pq\left(\begin{array}{cc}
            1&0\\0&-1
           \end{array}\right),\\ \\
Q_2=-\frac{1}{2}\left(\begin{array}{cc}
            0&p_x\\q_x&0
           \end{array}\right),  \qquad &
\cU_2=-\left(\begin{array}{cc}
            pq_x&0\\0&qp_x
           \end{array}\right),\\ \\
Q_3=\frac{1}{4}\left(\begin{array}{cc}
            0&-p_{xx}+4p^2q\\q_{xx}-q^2p&0
           \end{array}\right),  &
\cU_3=\frac{1}{2}\left(\!\begin{array}{cc}
            pq_{xx}-4p^2q^2&0\\0&4p^2q^2-qp_{xx}\!
           \end{array}\right),\ldots
\end{array} 
\]
and
\[
\begin{array}{l}
\rho_1=-2pq,\ \sigma_1^2=q_x p-p_x q,\  
\sigma_1^3=\frac{1}{2}(p_xq_x-pq_{xx}-qp_{xx})+6p^2q^2\, \ldots\, ,
\\ \\
  \rho_2=-qp_x,\ 
\sigma_2^2=\frac{1}{2}(p_xq_x-qp_{xx})+2p^2q^2 ,\
\sigma_2^3=\frac{1}{4}(p_{xx}q_x-qp_{xxx})+4pq^2p_x\, \ldots\, ,
\\ \\
\rho_3=2p^2q^2-\frac{1}{2}qp_{xx},\ 
\rho_4=pq(pq_x+4qp_x))-\frac{1}{4}qp_{xxx}\, \ldots\, .
\end{array}
\]

\subsection{Formal diagonalisation of the Darboux matrices $M_\alpha, N_h$.}
  
The diagonalising transformation $Q$ can be extended to the Darboux matrices 
$M_\alpha,M_\beta$ and $N_h$. 
We substitute $L=Q\cL Q^{-1}$ in  $\cS_\alpha (L)=M_\alpha 
LM_\alpha^{-1}$   to obtain
\begin{equation}\label{cM}
\cS_\alpha (\cL)=\cM_\alpha \cL 
\cM_\alpha^{-1}, \qquad \cM_\alpha=\cS_\alpha (Q)^{-1} M_\alpha Q.
\end{equation}
Similarly we obtain 
\[
 \cT_h (\cL)=\cN_h \cL 
\cN_h^{-1}, \qquad \cN_h=\cT_h (Q)^{-1} N Q.
\]
These equations can be rewritten in the form
\begin{eqnarray}\label{cMx}
 &&D_x(\cM_\alpha)=\cS_\alpha(\cL)\cM_\alpha-\cM_\alpha \cL\, ,\\ \label{cNx}
 &&D_x(\cN_h)=\cT_h(\cL)\cN_h-\cN_h \cL\, .
\end{eqnarray}

\begin{Pro}
 The coefficients $\cM_\alpha^k,\ \cN_h^k$ of the formal series
 \[
  \cM_\alpha=\cS_\alpha (Q)^{-1} M_\alpha Q=\lambda 
\cM_\alpha^{-1}+\cM_\alpha^{0}+\lambda ^{-1}
\cM_\alpha^{1}+\lambda ^{-2}
\cM_\alpha^{2}+\cdots
 \]
 and
  \[
  \cN_h=\cT_h (Q)^{-1} N Q=\lambda 
\cN_h^{-1}+\cN_h^{0}+\lambda ^{-1}
\cN_h^{1}+\lambda ^{-2}
\cN_h^{2}+\cdots 
 \]
are diagonal matrices.
\end{Pro}

{\bf Proof.} We prove it by induction. The leading coefficients 
$\cM_\alpha^{-1}=\cN_h{-1}$ are diagonal matrices with $(1,0)$ on the 
diagonal. Let us assume that the coefficients $\cM_\alpha^0,\ldots 
,\cM_\alpha^n$ are diagonal. Taking the coefficient $c_n$ at $\lambda^{-n}$ in
$D_x(\cM_\alpha)-\cS_\alpha(\cL)\cM_\alpha+\cM_\alpha \cL$ we obtain
\[
c_n=D_x(\cM_\alpha^n)-[J,\cM_\alpha^{n+1}]-\sum_{
k=-1}^n \cS_\alpha(\cU_{n-k})\cM_\alpha^k+\cM_\alpha^k \cU_{n-k}=0.
\]
Projection $\pi_\perp (c_n)=-[J,\cM_\alpha^{n+1}]=0$, which implies that 
$\cM_\alpha^{n+1}$ is diagonal. The proof is similar for the 
coefficients of $\cN_h$.\hfill $\Box$

Equation (\ref{cMx}) can be written in 
the form
\[
 D_x(\log \cM_\alpha)=(\cS_\alpha -I) \cL\, ,
\]
since all matrices in (\ref{cMx}) are diagonal.
Thus $\log \cM_\alpha$ is a generating function for local conservation laws for 
the differential-difference equation (\ref{meq}):
\[
 \log 
(\cM_\alpha)_{2,2}=\frac{r_\alpha^1}{\lambda}+\frac{r_\alpha^2}{\lambda^2}+\frac
{r_\alpha^3}{ \lambda^ 3} +\cdots
\]
\[
 D_x(r_\alpha^k)=\cS_\alpha (\rho_k)-\rho_k.
\]
It follows from 
(\ref{backlundtk}) and Proposition 1 that 
\[
 D_{t_k}(r_\alpha^n)=\cS_\alpha (\sigma^n_k)-\sigma^n_k.
\]
Moreover, it follows from (\ref{mamb}) that
\[
 \cS_\alpha(\cM_\beta)\cM_\alpha=\cS_\beta(\cM_\alpha)\cM_\beta .
\]
Therefore 
\[
 (\cS_\alpha-I)\log \cM_\beta= (\cS_\beta-I)\log \cM_\alpha
\]
and thus 
\[
 (\cS_\alpha-I)r^k_\beta= (\cS_\beta-I)r^k_\alpha,\qquad k=1,2,\ldots .
\]
Similarly, from (\ref{manh}) it follows that $
\cT_h(\cM_\alpha)\cN_h=\cS_\alpha(\cN_h)\cM_\alpha $ and thus 
\[
 (\cS_\alpha-I)r^k= (\cT_h-I)r^k_\alpha,\qquad \log 
(\cN_h)_{2,2}=\frac{r^1}{\lambda}+\frac{r^2}{\lambda^2}+\frac
{r^3}{ \lambda^ 3} +\cdots .
\]

{\bf Example.} Using equations (\ref{Meq}), (\ref{Neq}) 
corresponding to $\cS_\alpha L=\Ad_{M_\alpha}L, \cT_h L=\Ad_N L$  for
elimination of all $x$-derivatives we get
\[
 Q=I+\lambda^{-1}\left(\!\begin{array}{cc}
                        0&-p\\q&0
                       
\end{array}\!\right)+\lambda^{-2}\left(\!\begin{array}{cc}
                        0&-\alpha p-\cS_\alpha (p)+p^2\cS_\alpha 
(q)\\
\cS_\alpha^{-1}(q)+\alpha q-\cS_\alpha^{-1}(p)q^2&0 \end{array}\!\right)+\cdots=
\]
\[
I+\lambda^{-1}\left(\!\begin{array}{cc}
                        0&-p\\q&0
                       
\end{array}\!\right)+\lambda^{-2}\left(\!\begin{array}{cc}
                        0&hp\\-\cT_h^{-1}(hp^{-1})&0
 \end{array}\!\right)+\lambda^{-3}\left(\!\begin{array}{cc}
                        0&-h^2p-\cT_h(p)\\ 
\cT_h^{-1}(h^2p^{-1})-\cT_h^{-2}(p^{-1})&0
 \end{array}\!\!\right)+\cdots
\]
and
\[
 \cM_\alpha=\lambda\left(\!\begin{array}{cc}
                        1&0\\0&0
                       \end{array}\!\right)+
                       \left(\!\begin{array}{cc}
            p\cS_\alpha(q)-\alpha&0\\0& 1\end{array}\!\right)
                       +\lambda^{-1}\left(\!\begin{array}{cc}
                        pq&0\\0&-p\cS_\alpha(q)\end{array}\!\right)+\cdots
\]
\[
 \cN_h=\lambda\left(\!\begin{array}{cc}
                        1&0\\0&0
                       \end{array}\!\right)+
                       \left(\!\begin{array}{cc}
           h&0\\0& 0\end{array}\!\right)
                       +\lambda^{-1}\left(\!\begin{array}{cc}   
p\cT_h^{-1}(p^{-1})&0\\0&-1\end{array}\!\right)+
\lambda^{-2}\left(\!\begin{array}{cc}   
-p\cT_h^{-1}(hp^{-1})&0\\0&h\end{array}\!\right)+\cdots
\]
Thus
\begin{eqnarray}
&&\rho_1=-2 pq=-2\exp(\phi-\cT_h^{-1}\phi),\label{rho1} \\
&&\rho_2=-qp_x=-2 \left(\! \alpha
pq+\cS_\alpha(p)q-p^2q\cS_\alpha(q)\!\right)=2\exp(\phi-\cT_h^{-1}\phi)h
\label{rho2}
 \\
&&\sigma^1_2=pq_x-qp_x,\  \ 
\sigma_2^2=4p^2q^2+\frac{1}{2}(p_xq_x-qp_{xx})\label{sigma12},\\
&& r^1_\alpha=-p\cS_\alpha(q),\ \
r^2_\alpha=\frac{1}{2}p^2q^2-\alpha p\cS_\alpha(q)-\cS_\alpha(pq),\label{r12}\\
&&r^1=-h,\ \ r^2=\frac{1}{2}h^2-\exp(\cT_h(\phi)-\phi),\ \
r^3=-\frac{1}{3}h^3+\exp(\cT_h(\phi)-\phi)(h+\cT_h(h)),\ldots \nonumber
\end{eqnarray}
In applications to differential-difference equations one also need to eliminate 
$x$--derivatives from $\sigma^1_2,\sigma_2^2$ using  equations (\ref{Meq}), 
(\ref{Neq}).

\subsection{Diagonalisation of adjacent Lax structure}

It is easy to justify that the transformation (\ref{Q}), which formally 
diagonalises 
the Lax operator $L$ (Lemma 1) and operators $A^k$, also diagonalises 
the operators $B_\alpha^k$ (\ref{Balphak}) associated with the 
adjacent Lax structure. For example, 
\[
 Q^{-1}B_\alpha^1 Q=\frac{\lambda^{-1}}{1+p_{-1}q_1}
\left(\!\begin{array}{cc} p_{-1}q_1&0\\0&1 \end{array}\!\right)+
\frac{\lambda^{-2}}{1+p_{-1}q_1} \left(\!\begin{array}{cc} p_{-1}(q+\alpha 
q_1)&0\\0&\alpha-pq_1
\end{array}\!\right)+\cdots. \]
Thus, the coefficients $\hat{\sigma}^k_\alpha$ 
\begin{eqnarray*}
&& \hat{\sigma}^1_\alpha =\frac{1}{1+p_{-1}q_1}=1-\sqrt{1+4p_y q_y},\\
&& \hat{\sigma}^2_\alpha =\frac{\alpha+pq_1}{1+p_{-1}q_1}=\alpha(1-\sqrt{1+4p_y 
q_y})+pq_y,\ \ldots
\end{eqnarray*}
in the expansion 
\[
 \left(\! Q^{-1}B_\alpha^1 Q\right)_{2,2}=\hat{\sigma}^1_\alpha 
\lambda^{-1}+\hat{\sigma}^2_\alpha \lambda^{-2}+\hat{\sigma}^3_\alpha 
\lambda^{-3}+\cdots 
\]
are fluxes for the local conservation laws of (\ref{dH})
\[
 D_y(r_\alpha^1)=(\cS_\alpha-1)\hat{\sigma}^1_\alpha, \qquad 
D_y(r_\alpha^2)=(\cS_\alpha-1)\hat{\sigma}^2_\alpha,\ldots 
\]
andfor (\ref{pqxix})
\[
 D_y(\rho_1)=D_x(\hat{\sigma}^1_\alpha), \qquad 
D_y(\rho_2)=D_x(\hat{\sigma}^2_\alpha),\ldots
\]
where $r_\alpha^1,r_\alpha^2$ and $\rho_1,\rho_1$ are given in (\ref{r12}) and 
(\ref{rho1}),(\ref{rho2}) respectively.

Lax operator $B_\alpha^1$ has a pole at $\lambda=\alpha$ and we can diagonalise 
it around this pole. It is convenient to introduce a local parameter 
$\mu=(\lambda-\alpha)^{-1}$ and diagonalise the coefficient at the pole by the 
gauge transformation
\begin{equation}
 \hat{B}_\alpha^1=T_0^{-1}B_\alpha^1 T_0=D_y-\mu J_1+\hat{W},
 \label{Bhat}
\end{equation}
where
\[
 T_0=\left(\begin{array}{cr}
    p_{-1}&-1\\1&q_1
     \end{array}\right),\quad J_1=\frac{1}{2}(I+J)=\left(\begin{array}{cc}
                                            1&0\\0&0
                                           \end{array}\right),\
\hat{W}=\left(\begin{array}{cc}
      -q_1 p_{-1,y}&-q_{1,y}\\p_{-1,y}&p_{-1}q_{1,y}
                                           \end{array}\right)    .               
 \]
\begin{Pro}\label{propT}
 Transformation $T^{-1}\hat{B}T:=\cB$ brings operator $\hat{B}$ (\ref{Bhat}) to 
a diagonal form
 \[
  \cB=D_y-\mu J_1-\cW_0-\mu^{-1}\cW_1-\mu^{-2}\cW_2+\cdots ,\qquad
\cW_k=\pi_\parallel  (\cW_k),
 \]
Where
\[
 \cW_0=\pi_\parallel (\hat{W}),\qquad \cW_k=\pi_\perp (\hat{W})T_k
\]
\[
 T=I+\mu^{-1}T_1+\mu^{-2}T_2+\cdots
\]
and off--diagonal coefficients $T_k$ can be found recursively
\begin{eqnarray*}
 &&T_1=-\frac{1}{2}\ad _J (\hat{W}),\\
 &&T_{k+1}=\frac{1}{2}\ad _J \left(T_{k,y}-\pi_\parallel 
(\hat{W})T_k+T_k\pi_\parallel (\hat{W})+\sum_{s=1}^{k-1}T_{k-s}\pi_\perp 
(\hat{W})T_s\right).
\end{eqnarray*}
\end{Pro}
 We omit the proof since it is very similar to the proof of Lemma 1.
 
The same transformation $ {\cL}=(T_0T)^{-1}LT_0 T$ brings the Lax operator 
$L$ to a diagonal form (this diagonalisation is different from the one given in 
Lemma 1). The coefficients $\varrho_k=(\cW_k)_{2,2}$ of the expansion
\[
 \varrho_0=-\frac{p_{-1}q_{1,y}}{1+p_{-1}q_1},\qquad 
\varrho_1=-\frac{p_{-1,y}q_{1,y}}{(1+p_{-1}q_1)^2},\ldots
\]
are densities of the conservation laws for the Heisenberg hierarchy (\ref{1H}), 
 principal chiral field model (\ref{chiralfield}) and system (\ref{pqxix}). 
 
It can be easily shown that transformation $\cM 
_\alpha=\cS_\alpha(T_0T)^{-1}M_\alpha T_0T$ brings the Darboux matrix 
$M_\alpha$ in a diagonal form $ \cM_\alpha$. To apply the transformation to 
$M_\alpha$ we need to eliminate the $y$--derivatives from the coefficients 
$T_k$ 
using equation (\ref{dH}).

There is a direct way  to diagonalise the 
Darboux matrix $M_\alpha$. Matrix $M_\alpha$ has two points on the Riemann 
sphere, where the leading 
coefficient (in the local parameter) is singular. Indeed, at $\lambda=\infty$ 
and  $\lambda=\alpha$ the leading coefficients are
\[
\lambda\left(\begin{array}{cl}
1& 0\\0 &0      ×
       \end{array}\right)\quad {\rm and}\quad 
\left(\begin{array}{cl}
pq_1& p\\q_1 &1      ×
       \end{array}\right)
\]
respectively.
Let us diagonalise the Darboux matrix at 
$\lambda=\alpha$ without using the result  Proposition \ref{propT}. Namely, we 
can find coefficients of a formal series 
\[
 T=I+\mu^{-1}T_1+\mu^{-2}T_2+\mu^{-3}T_3+\cdots ,\qquad T_k=\pi_\perp (T_k)
\]
such that the coefficients $\tilde{\cM}_k$ in 
\begin{equation}
\label{cMtilde}
{ \tilde{\cM}}_\alpha=\cS_\alpha(T)^{-1}\tilde{M}_\alpha 
T= { \tilde{\cM}}_0+\mu^{-1} { \tilde{\cM}}_{1}+\mu^{-2} { 
\tilde{\cM}}_{2}+\cdots, \qquad  { \tilde{\cM}}_{k}=\pi_\parallel( { 
\tilde{\cM}}_{k})
\end{equation}
where 
\[
 \tilde{M}_\alpha=\cS_\alpha(T_0)^{-1}M_\alpha 
T_0=\tilde{M}_0+\mu^{-1}\tilde{M}_1,\]
\[ \tilde{M}_0=\left(\begin{array}{cc}
           1+p_{-1}q_1&0\\0&0   \end{array}\right),\ 
\tilde{M}_1=\frac{1}{1+pq_2}\left(\begin{array}{cc}
           p_{-1}q_2&-q_2\\-p_{-1}&1   \end{array}\right) .                      
     \]

\begin{Pro}\label{prop4}T
 The coefficients $T_k,\ T_k=\pi_\perp (T_k)$ such that  the coefficients  
$\tilde{\cM}_{0}=\tilde{\cM}_0,\ \tilde{\cM}_{k+1}=\pi_\perp 
(\tilde{M}_1)T_k$ are diagonal can be found recursively
\begin{eqnarray*}
&& T_1=\frac{1}{×(1+p_{-1}q_{1})(1+pq_{2})} \left(\begin{array}{ll}
0&q_2\\-p_{-1}&0     \end{array}\right),\\                                      
 && \tilde{\cM}_0T_{k+1}-\cS_\alpha (T_{k+1}) 
\tilde{\cM}_0+\pi_\parallel(\tilde{M}_{1})T_k-\sum_{s=1}^{k} 
\cS_\alpha(T_s)\pi_\perp 
(\tilde{M}_1)T_{k-s}=0.                                             
\end{eqnarray*}

\end{Pro}

The proof is straightforward. What is important here is  that 
in the recursion we 
do not 
need to solve difference equations since ${\rm rank} \tilde{\cM}_0=1$ and 
${\rm Ker} \tilde{\cM}_0\bigoplus{\rm Im} \tilde{\cM}_0=\C^2$. Therefore all 
entries of $T_k$ and  $\tilde{\cM}_k$ are elements of the difference ring 
$[\C;p,q,(1+p_{-1}q_1)^{1};\cS_\alpha]$, i.e difference polynomials of 
variables 
$ p,q,(1+p_{-1}q_1)^{-1}$ and their $\cS_\alpha^k,\ k\in \Z$ shifts with 
complex 
coefficients.

It follows from Proposition \ref{prop4} that
\[
 \tilde{\cM}_\alpha=\left(\!\begin{array}{cc}
           1+p_{-1}q_1&0\\0&0   \end{array}\!\right)+\frac{\mu^{-1}}{1+pq_2}
           \left(\!\begin{array}{cc}
           p_{-1}q_2&0\\0&1   \end{array}\!\right)+\frac{\mu^{-2}}{1+p_{-1}q_1}
           \left(\!\begin{array}{cc}
           \frac{p_{-2}q_2}{1+p_{-1}q}&0\\0&-\frac{p_{-1}q_2}{1+pq_2}   
\end{array}\!\right)+\cdots
\]
and thus the coefficients $\tilde{r_k}_\alpha$ 
\[
 \tilde{r^0}_\alpha=-\log(1+pq_{2}),\quad 
\tilde{r^1}_\alpha=\frac{p_{-1}q_2}{(1+pq_{2})(1+p_{-1}q_1)}, \ldots
\]
in the expansion of 
$(\tilde{\cM}_\alpha)_{2,2}=-\log 
(\mu)+\tilde{r^0}_\alpha+\mu^{-1}\tilde{r^1}_\alpha+\cdots$ are new densities 
of local conservation laws for differential difference  equations (\ref{meq}), 
(\ref{dH}) and partial difference equations (\ref{dnls}), (\ref{dtoda}).

It is obvious that the transformation constructed in Proposition \ref{propT} 
and in Proposition \ref{prop4} coincide modulo equation (\ref{dH}) and these 
two approaches are equivalent.

\subsection{Summary}

In this paper we have  presented the concept of Lax-Darboux scheme and 
illustrated it on the example of the
NLS equation. From the differential-difference algebra point of view 
the scheme can be described as follows. The base object is a 
differential-difference ring polynomials
$\cR=[\C;\bu;D_{x_1},D_{x_2},\ldots ;\cS_1,\cS_2,\ldots]$  of a (vector) 
variable $\bu={u^1,\ldots , u^M}$, its derivatives and shifts 
$D_{x_1}^{n_1}\cdots D_{x_m}^{n_m}\cS_1^{s_1}\ldots \cS_{p}^{s_p}\bu $, equipped 
with a  set of commuting derivations $D_{x_k},\ 
k=1,2,\ldots$ 
and commuting automorphisms $\cS_i,\ i=1,2\ldots$. To each $D_{x_k}$ we 
associate a Lax operator of the form $L^{k}=D_{x_k}-U^k$, where $U^k$ is  
$N\times N$ matrix with entries belonging to $\cR(\lambda)$, i.e. are rational 
functions of a spectral parameter $\lambda$ with coefficients from $\cR$. With 
each automorphism $\cS_i$ we associate a Darboux $N\times N$ matrix $M^i$ with 
entries from $\cR(\lambda)$. Then the system of Lax-Darboux equations we 
identify with the differential-difference ideal  
\[
 \cI=\langle [L^i,L^j],\ \cS_i(L^j)-\Ad_{M^i}(L^{j}),\ 
\cS_i(M^j)M^i-\cS_j(M^j)M^j\rangle\ \subset\cR
\]
and consider a quotient ring $\cR_\cI=\cR\diagup \cI$. In this setup a
statement  that two expressions are equal modulo equations simply means that 
these two expressions are  equal as elements of $\cR_\cI$. 

We can formally diagonalise (or bring to a 
block-diagonal form) simultaneously all matrices $U^i,M^j$ near singular 
points in $\lambda$ and generate infinite sequences $\rho^i_k,r^j_k,\ 
k=1,2,\ldots $ such that in  $\cR_\cI$ they satisfy relations
\begin{eqnarray*}
  &&D_{x_n}\rho_k^j= D_{x_j}\rho_k^n,\\&&  D_{x_n}r_k^j=(\cS_j-1)\rho_k^n,\\&&
(\cS_i-1)r_k^j=(\cS_j-1)r_k^i.
\end{eqnarray*}
These relations can be regarded as a sequences of local conservation laws for 
partial differential, differential difference and partial difference  equations.
The Lax-Darboux scheme can be generalised to the case when the derivations are 
not commuting, but such generalisatins are beyond of the scope of this paper.

In the case of the NLS equation the elements of the  
Lax-Darboux scheme are:
\begin{itemize}
 \item The Lax structure, i.e. Lax operators $L,A^k$ such that the 
commutativity conditions
$[L,A^k]=[A^k,A^p]=0$ are equivalent to a system of integrable partial 
differential equations and its symmetries (\ref{LAlaxstr}).
\item Darboux transformations  $\cS_\alpha,\cT_h$ with Darboux matrices 
$M_\alpha$ (\ref{Malpha}) and $N_h$ (\ref{Nh}) 
respectively. The compatibility conditions 
$\cS_\alpha L=\Ad_{M^\alpha}L,\ \cT_h L=\Ad_{N_h}L$ and
$\cS_\alpha A^k=\Ad_{M^\alpha}A^k,\ \cT_h A^k=\Ad_{N_h}A^k$  result in 
B\"acklund 
transformations of the above 
integrable system and its symmetries (\ref{meq}), (\ref{todamap}). B\"acklund 
transformations also 
can be regarded as integrable 
differential-difference equations in their own right.
\item The Bianchi lattices, which follow from the commutativity conditions 
for 
pairs of Darboux transformations  result in   integrable 
systems 
of partial difference equations (\ref{dnls}), (\ref{dtoda}). The mentioned 
above differential-difference 
equations (\ref{meq}), (\ref{todamap}) are symmetries of these systems.
\item There is an  adjacent Lax structure  (corresponding to 
operators  $B_\alpha^k$ 
(\ref{Balphak})) sharing the same Darboux matrix $M_\alpha$ and resulting in 
the 
differential-difference integrable system (\ref{dH}). The commutativity 
condition $[B_\alpha^1,B_\alpha^k]=0$ results in the hierarchy of the 
Heisenberg 
model (\ref{1H}). The commutativity condition 
$[B_\alpha^1,B_\beta^1]=0$ is equivalent to the principal chiral field model 
(\ref{chiralfield}), so that the hierarchy of the Heisenberg equation is a 
hierarchy of symmetries for (\ref{chiralfield}). Equation 
$[L,B_\alpha^k]=0$ provide us with a hierarchy of symmetries for system 
(\ref{pqxix}). Integrable differential-difference systems of equations arising 
from the conditions 
$\cS_\alpha(B_\alpha^k)=\Ad_{M_\alpha}(B_\alpha^k)$ (such as  (\ref{dH}) and  
(\ref{dH2})) are B\"acklund transformations for the above listed hierarchies 
and symmetries for differential-difference equations (\ref{meq}), 
(\ref{todamap}) and partial difference equations (\ref{dnls}), (\ref{dtoda}).
\end{itemize}

We have shown that there is a formal diagonalistaion of the Lax-Darboux scheme, 
i.e. a transformation (in the form of a formal series in the spectral 
parameter) which
diagonalises simultaneously the Lax structure, associated Darboux 
transformations and adjacent Lax structures. The diagonalised Lax (and adjacent 
Lax) operators and logarithms of the diagonalised Darboux matrices are 
generating functions of local conservation laws (both the densities and fluxes) 
for related partial differential, differential-difference and partial 
difference equations, which  are neatly related to each other. Moreover, there 
may 
exist several different diagonalisations, which lead to adjacent hierarchies of 
local conservation laws for equations corresponding to the Lax-Darboux scheme.  

\section*{Acknowledgements} I would like to thank participants of my lecture 
courses, seminars 
and conferences, where several parts of this study were presented, for 
their useful comments. In particular I am grateful to G.Berkeley, 
R..N.Garifullin, 
S.Konstantinou-Rizos, V.V.Sokolov, 
R.I.Yamilov and J.P.Wang for discussions and advice. I am gratefully 
acknowledge support from the Leverhulme Trust.
 

\end{document}